# Periodic Behavior of Topology in Graphene with Nanohole Array

Yong-Cheng Jiang[1], Xing-Xiang Wang[1], and Xiao Hu[1,2*]

[1] *Institute for Quantum Science and Technology, and Department of Physics, Shanghai University, Shanghai 200444, China*

[2] *School of Engineering, Institute of Science Tokyo, Tokyo 152–8552, Japan*

We derive a way to diagnose band topology for graphene with triangular and/or honeycomb array of nanoholes directly from the lattice constant of superstructure $m\sqrt{3} \times m\sqrt{3}$ with integer $m$. Taking into account the $C_{6v}$ crystalline symmetry respected by nanoholes and their array, we demonstrate that nontrivial topology appears periodically with $m$ with period two (six) for triangular (honeycomb) array. These behaviors are verified by Wyckoff positions of Wannier centers and parity index of valence bands at high-symmetry points in Brillouin zone. The results provide a convenient guide for material design of topological electronic states based on graphene derivatives.

*Introduction*. With the increasing density of transistors in integrated circuits, quantum tunneling and thermal dissipation have become increasingly severe, indicating that Moore's law is approaching to its limit. Dissipationless electric currents in quantum Hall effect exhibit the potential to address these issues,[1–6)] while quantum spin Hall effect opens a new avenue for spintronics.[7–11)] Understanding Berry curvature and topological properties of electronic states[12)] provides new opportunities for exploration of topological materials and exploitation of quantum phenomena. Despite significant progress in recent years,[13,14)] the realization of quantized transport is still limited to low temperatures and small scales. Therefore, the search for topological materials with large energy gaps has become an important topic.

By deforming honeycomb lattice while preserving $C_{6v}$ crystalline symmetry, one can realize topological crystalline states[15–19)] with a large energy gap in Dirac dispersions of graphene,[20)] without resorting to strong spin-orbit coupling. The nontrivial topology is captured by an effective $\mathbf{k} \cdot \mathbf{p}$ Hamiltonian at $\Gamma$ point with the same form as the BHZ model for quantum spin Hall effect,[9)] and can be further understood in terms of Euler class when $C_2T$ symmetry is exploited over the entire

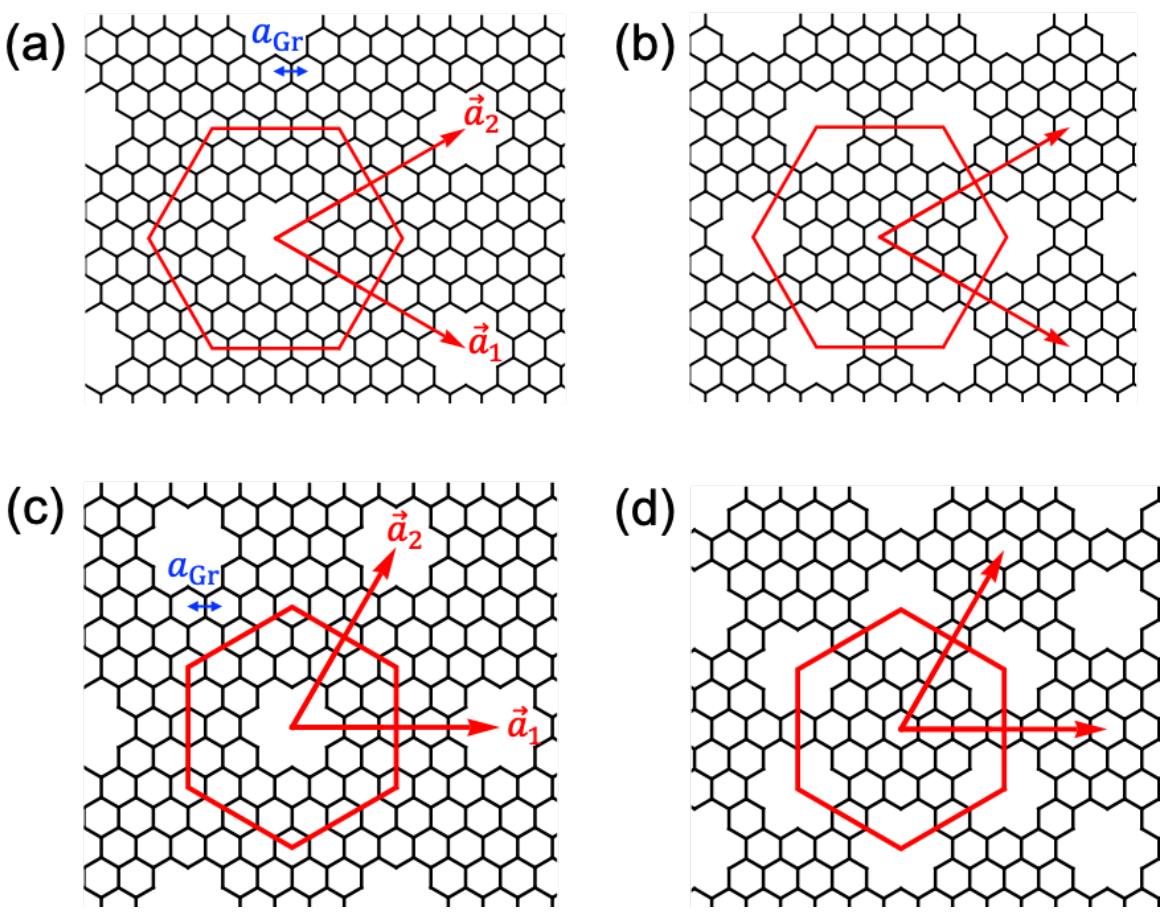

**Fig. 1**. (Color online) Schematic structures for graphene with (a) $4\sqrt{3} \times 4\sqrt{3}$ triangular, (b) $4\sqrt{3} \times 4\sqrt{3}$ honeycomb, (c) $6 \times 6$ triangular, and (d) $6 \times 6$ honeycomb nanohole array. $\vec{a}_{1,2}$ are unit vectors of superstructured graphene, and $a_{\rm Gr}$ is the lattice constant of pristine graphene. The hexagonal unit cells are chosen for topological characterization.

BZ.[21–24] The topological edge states have been realized in molecular graphene[25,26] following the theoretical proposal,[27] and Kekulé distortions can be induced in graphene by noble-gas deposition[28–31] and alkali-metal intercalation.[32] State-of-the-art quantum Monte Carlo calculations predict that graphene under isotropical tensile strain experiences a structure transition to a stable Kekulé-like dimerized phase with a topological energy gap up to 1 eV.[33] As an alternative approach, topological crystalline states can be realized in graphene by introducing regular arrays of nanoholes,[34–39] which may be fabricated using bottom-up synthesis methods.[40,41]

Recently it has been formulated[42–46] that, distinctly from the trivial atomic limit, nontrivial topology can be characterized by obstruction to localization, or localization at non-atomic Wyckoff positions of Wannier functions[47] associated with relevant bands. With information of Wannier functions, topological materials have been diagnosed comprehensively in material database[48–51] using topological quantum chemistry[45,46] and symmetry indicators.[52] The positions of Wannier centers provide essential information for characterization of band topology.

In this Letter, we systematically investigate the topology in graphene with $C_{6v}$-symmetric nanohole arrays. We focus on the $m\sqrt{3} \times m\sqrt{3}$ and $3m \times 3m$ superstructures, where the energy gap is opened in double Dirac cones at the Fermi level by the intervalley interaction due to the Brillouin zone folding. Topology diagnose has been performed based on analysis on Wyckoff positions of Wannier centers and the parity index of valence bands at high-symmetry points in BZ calculated in terms of tight-binding Hamiltonian. Tables of topology have been established for triangular and honeycomb arrays of nanohole superstructures, where periodic behaviors with respect to lattice constant of superstructure are clearly observed, which are expected useful for designing topological materials.

*Tight-binding model*. Graphene with nanohole array is formed by removing C atom clusters periodically in pristine graphene, as depicted in Fig. 1. We start with nanoholes composed of six C atoms, and focus on triangular or honeycomb arrays, where the superstructured graphene enjoys $C_{6v}$ crystalline symmetry. We denote pristine graphene as the $1 \times 1$ structure for its two unit vectors with the lattice constant $a_{\mathrm{Gr}}$, so that the $L \times L$ superstructure represents that the lattice constants of its unit vectors are $|\vec{a}_{1,2}|/a_{\mathrm{Gr}} = L$.

In order to describe electronic properties of graphene with nanohole array, we adopt the following tight-binding (TB) Hamiltonian for $\pi$ electrons:

$$H = -t \sum_{\langle i,j \rangle} c_i^{\dagger} c_j + \mathrm{H.c.}, \qquad (1)$$

where only a uniform nearest-neighbor hopping energy $t$ is considered,[53] and the C atoms at the perimeters of nanoholes are hydrogenated. For the band structure of pristine graphene, two Dirac cones are located at K and K' points in the Brillouin zone (BZ), respectively.[20] When an array of nanoholes is introduced such that the BZ folding of the superstructure maps these two momenta onto the $\Gamma$ point, an energy gap is opened at the Fermi level of half filling, where intervalley interactions play an important role.[54,55] This situation occurs exclusively in $m\sqrt{3} \times m\sqrt{3}$ and $3m \times 3m$ superstructures, with integer $m$, which is focused in this work.

*Triangular array of nanoholes*. With the TB Hamiltonian (1) we calculate the band structures for graphene with a triangular array of nanoholes. The topology of the energy gap at the Fermi level can be characterized by using parity index,[34] defined as the number of valence bands (VBs) with even parity against the $C_2$ rotation with respect to the center of the hexagonal unit cell (see Fig. 1), at the high-symmetry momenta $\Gamma$ and M points. The balance/imbalance of parity indices between $\Gamma$ and M points indicates that the system is trivial/topological. Based on this criterion we assess the topology for $m\sqrt{3} \times m\sqrt{3}$ and $3m \times 3m$ triangular superstructures as summarized in Tables Table I and Table II (see Tables S1 and S2 in Supplemental Material for details). A period of two is clearly observed: the triangularly superstructured graphene is topological/trivial for even/odd $m$.

Table I. Periodic table of topology in graphene with $m\sqrt{3}\times m\sqrt{3}$ nanohole array. ✓/✗: topological/trivial; $N_{\mathrm{VB}}^{\triangle}/N_{\mathrm{VB}}^{\hexagon}$ : number of valence bands in triangular/honeycomb nanohole-array. N/A: not applicable as the structure does not exist.

| $m$ | Triangular | $N_{\mathrm{VB}}^{\triangle}$ (mod 6) | Honeycomb | $N_{\mathrm{VB}}^{\hexagon}$ (mod 6) |
|---|---|---|---|---|
| 2 | ✓ | 3 | N/A | N/A |
| 3 | ✗ | 0 | ✗ | 3 |
| 4 | ✓ | 3 | ✗ | 0 |
| 5 | ✗ | 0 | ✓ | 3 |
| 6 | ✓ | 3 | ✓ | 0 |
| 7 | ✗ | 0 | ✓ | 3 |
| 8 | ✓ | 3 | ✗ | 0 |
| 9 | ✗ | 0 | ✗ | 3 |
| 10 | ✓ | 3 | ✗ | 0 |
| 11 | ✗ | 0 | ✓ | 3 |

Table II. Periodic table of topology in graphene with $3m\times 3m$ nanohole array.

| $m$ | Triangular | $N_{\mathrm{VB}}^{\triangle}$ (mod 6) | Honeycomb | $N_{\mathrm{VB}}^{\hexagon}$ (mod 6) |
|---|---|---|---|---|
| 1 | ✗ | 0 | N/A | N/A |
| 2 | ✓ | 3 | ✓ | 0 |
| 3 | ✗ | 0 | ✗ | 3 |
| 4 | ✓ | 3 | ✓ | 0 |

In order to understand this periodic behavior of topology with respect to the superstructure lattice constant, we count the number of VBs $N_{\mathrm{VB}}^{\triangle}$ at half filling. For the $m\sqrt{3}\times m\sqrt{3}$ and $3m\times 3m$ triangular nanohole arrays with six C atoms removed, it is straightforward to obtain $N_{\mathrm{VB}}^{\triangle}=3(m^2-1)$ and $3(3m^2-1)$, respectively. For both series, the number of VBs modulo 6, $N_{\mathrm{VB}}^{\triangle}$ (mod 6), is 0/3 for $m$ odd/even. Intriguingly, we find that for superstructures with $N_{\mathrm{VB}}^{\triangle}\ (\mathrm{mod}\ 6)=0/3$ the energy gap at half filling is trivial/topological. The one-to-one correspondence between topology and $N_{\mathrm{VB}}^{\triangle}$ (mod 6) can be explained by analyzing the positions of Wannier centers. Since the structures are characterized by $C_{6v}$ crystalline symmetry, the

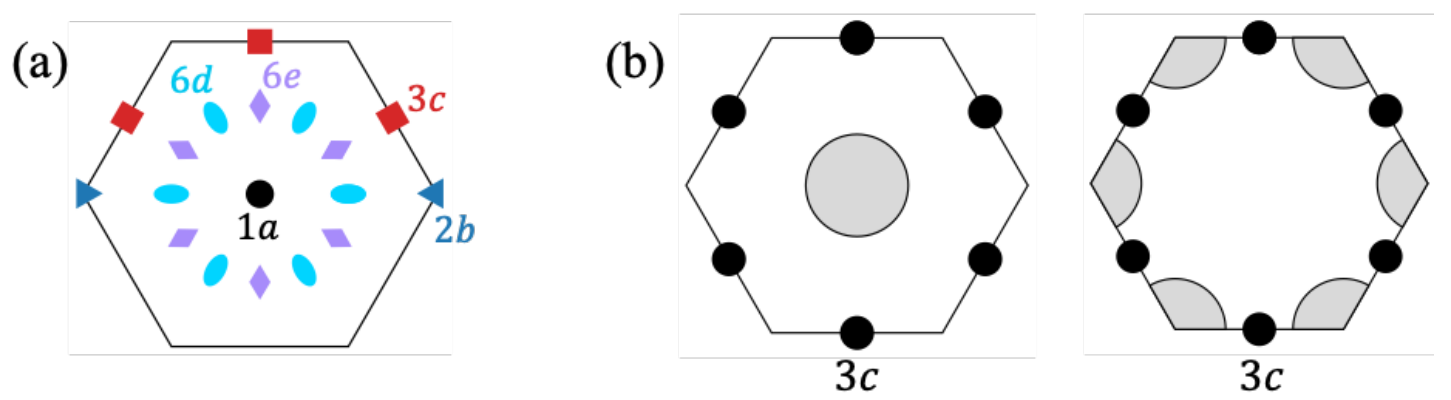

**Fig. 2.** (Color online) (a) Wyckoff positions for a $C_{6v}$-symmetric unit cell. (b) Wannier centers (black dots) located at $3c$ Wyckoff positions for three valence bands in graphene with triangular (left) and honeycomb (right) arrays of nanoholes (grey areas).

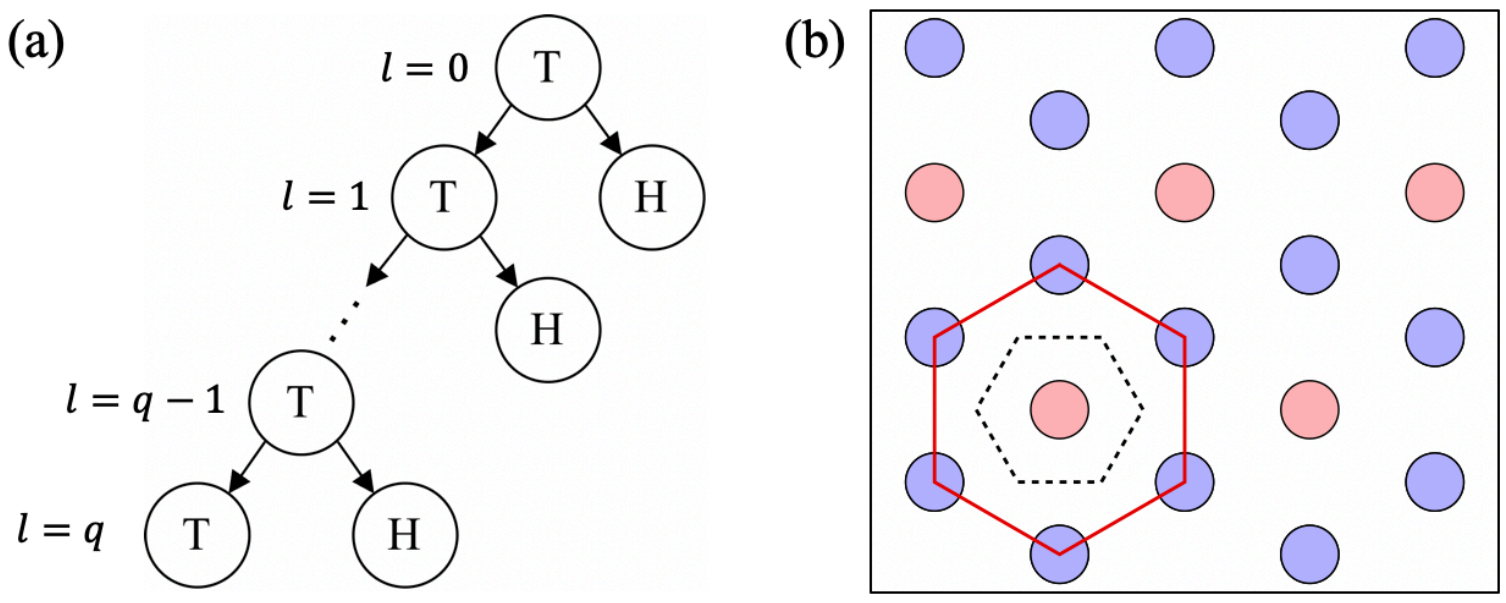

**Fig. 3.** (Color online) (a) Folding tree for superstructures with lattice constants $L = (3p \pm 1) \times (\sqrt{3})^q$. $l$: folding times; T/H: triangular/honeycomb array. (b) Schematic structures for $l$ triangular (both red and blue), $l+1$ triangular (red), and $l+1$ honeycomb (blue) arrays, with nanoholes denoted by circles. The dashed (solid) hexagon represents the unit cell of structure for $l$ triangular ($l+1$, both triangular and honeycomb) array.

Wannier centers of any set of six entangled bands locate at the $6d$ or $6e$ Wyckoff positions depicted in Fig. 2(a), which results in trivial band topology. For the case $N_{\mathrm{VB}}^{\triangle}$ (mod 6) = 3, the Wannier centers associated with the leftover three VBs near the Fermi level can only occupy the three $3c$ Wyckoff positions depicted in Fig. 2(b), since $1a$ Wyckoff positions are punctured by nanoholes, which renders a nontrivial band topology.[45,46,56]

The above results reveal that the two ways to assess the band topology for superstructures characterized by $C_{6v}$ crystalline symmetry, one based on imbalance of parity index of the VBs and the other based on nonzero value of number of VBs modulo 6, are intimately related to each other. The latter one is more useful for material design since one only has to count the number of sites.

*Honeycomb array of nanoholes*. Following the above discussions on triangular arrays of nanoholes, we first count $N_{\mathrm{VB}}^{\hexagon}$ (mod 6). It is found that a honeycomb array has three fewer VBs than the corresponding triangular one with the same lattice constant, which implies distinct topology in the two superstructures. Analyzing the parity index of VBs, we observe that this rule is satisfied except for $3m\sqrt{3} \times 3m\sqrt{3}$ and/or $3m \times 3m$, as summarized in Tables Table I and Table II (see Tables S1 and S2 in Supplemental Material[56] for details).

The topology in honeycomb arrays with lattice constants $3m\sqrt{3} \times 3m\sqrt{3}$ and $3m \times 3m$ should be diagnosed more carefully as detailed in what follows. First, we observe that these superstructures can be described by $L \times L$ with $L = (3p \pm 1) \times (\sqrt{3})^q$, $p \geq 0$, $q \geq 2$with integers $p$, $q$; for $p = 0$, + sign should be taken (Note, for $q = 1$ a honeycomb array exhibits distinct topology to the corresponding triangular array as addressed above). The VBs of a honeycomb array are obtained by first $(q-1)$-times $\sqrt{3}\times\sqrt{3}$ BZ folding in triangular form from the pristine graphene and finally one $\sqrt{3}\times\sqrt{3}$ BZ folding in a honeycomb form (see Fig. 3). The first $\sqrt{3}\times\sqrt{3}$ BZ folding in triangular form generates a primary energy gap at the half-filling Fermi level from double Dirac cones, while the following $\sqrt{3}\times\sqrt{3}$ BZ foldings merely rescale the energy gap without gap closing, since potentials with larger lattice constants induce weaker intervalley couplings. Therefore, the topology of honeycomb array of $L=(3p\pm1)\times(\sqrt{3})^{q}, q\geq 2$ inherits directly from the triangular array of $L_1=(3p\pm 1)\times\sqrt{3}$. For honeycomb array with $L=3\sqrt{3}$, namely $p=0,\ q=3$, the topology inherits from the triangular array $3\times 3$, which is trivial as given in Table II. It is then straightforward to confirm that, for honeycomb arrays, a period of six in topology with respect to the superstructure lattice constant exists for $m\sqrt{3}\times m\sqrt{3}$, while period of two remains unchanged for $3m\times 3m$, as summarized in Tables Table I and Table II. It is easy to check that, for triangular arrays of nanoholes, $N_{\mathrm{VB}}^{\triangle}\ (\mathrm{mod}\ 6)$ and the above band folding result in the same topology as it should be (see Supplementary Note I).[56)]

*Discussions.* Nanoholes larger than six C atoms can be used to compose regular superstructures. We find that topology assessment based on nonzero value of number of VBs modulo 6 works perfectly for triangular array (see Table S3 in Supplementary Material[56)] for details). For honeycomb arrays with narrow neck width between large holes, exceptional cases may occur when both $1a$ and $3c$ Wyckoff positions can host three Wannier centers simultaneously (in total six), where analysis on parity index is required to diagnose band topology (see Table S4 in Supplementary Material[56)] for details).

A closely related system, graphene with patterned hydrogen adatoms, was analyzed before, where nanoholes are formed effectively by eliminating $\pi$ electrons of carbon atoms.[57)] The imbalances of parity indices between $\Gamma$ and M points have been analyzed by both density-functional-theory calculations and TB modelling, which give the same results as expected. Therefore, the present results are considered applicable to graphene with patterned hydrogen adatoms.

In the present work, triangular and/or honeycomb array of nanoholes are introduced in the pristine graphene with $C_{6v}$ symmetry respected, which bring $\mathrm{K}_{\mathrm{Gr}}$ and $\mathrm{K}'_{\mathrm{Gr}}$ points of the pristine graphene to the $\Gamma$ point of the small, folded BZ of the superstructure. Removing carbon atoms to form nanoholes is physically equivalent to introduce strong, periodic potential, and thus open a gap in the otherwise Dirac cones at the $\Gamma$ point (see Supplementary Note II[56] for details). This situation is similar to the O-shaped Kekulé with $C_{6v}$ symmetry, where intervalley coupling is induced by non-uniform hopping integrals[16,17,58]. For other Kekulé-type graphene superstructures, such as Y-shaped Kekulé,[59,60] $\mathrm{K_{Gr}}$ and $\mathrm{K'_{Gr}}$ points are generally mapped to different momenta. In the latter case, no intervalley coupling occurs, and two Dirac cones locating at different momenta remain intact even though non-uniform hopping integrals exist.

Imperfections and various disorder may affect properties of the present system. However, as long as they do not close the band gap, the topology of the structures including the positions of Wannier centers remains unchanged. The size of energy gap in graphene with nanohole array can be estimated with a relation $E_g \approx 25\ \mathrm{eV} \times N_{\mathrm{removed}}^{1/2}/N_{\mathrm{total}} = 25\ \mathrm{eV} \times \sqrt{6}/(2L^2 - 6) \approx 30.6/(L^2 - 3)\ \mathrm{eV}$, where $N_{\mathrm{removed}}$ and $N_{\mathrm{total}}$ are the number of removed carbon sites and the number of total carbon sites in a unit cell.[61] Energy gaps of systems with unit cell of several nanometers in linear size are around sub-eV, which sets a limit of the strength of disorder.

*Summary*. We present a periodic table for the band topology of graphene derivatives where triangular and/or honeycomb arrays of nanoholes are punctured in pristine graphene, which can be used to perform topology diagnose sorely by the lattice constant. The correspondence between the band topology and the purely real-space information is elucidated in terms of positions of Wannier centers associated with valence bands, where the $C_{6v}$ crystalline symmetry and locations of nanoholes play important roles. Our work unveils a mechanism for realizing topological crystalline states in graphene derivatives through real-space engineering, enabling efficient discovery and design of topological materials and their realization using cutting-edge nanofabrication techniques.

**Acknowledgments**

The authors are grateful to T. Kariyado for valuable discussions. This work is supported by Shanghai Science and Technology Innovation Action Plan (No. 24LZ1400800), and partially supported by CREST, JST (Core Research for Evolutionary Science and Technology, Japan Science and Technology Agency) (Grant Number JPMJCR18T4).

*Corresponding author. Email: hu_xiao@shu.edu.cn

# Supplementary Material

## Periodic Behavior of Topology in Graphene with Nanohole Array

Yong-Cheng Jiang[1], Xing-Xiang Wang[1] and Xiao Hu[1,2*]

[1]*Institute for Quantum Science and Technology, and Department of Physics,*
*Shanghai University, Shanghai 200444, China*
[2]*School of Engineering, Institute of Science Tokyo, Tokyo 152–8552, Japan*

*Corresponding author. Email: hu_xiao@shu.edu.cn

**Supplementary Note I**: The number of VBs at half filling for a triangular array with lattice constant $L = (3p \pm 1) \times \left(\sqrt{3}\right)^q, q \geq 2$ with integers $p,\ q$ is

$$N_{\mathrm{VB}}^{\triangle}(L) = L^2 - 3 = 3^q(3p \pm 1)^2 - 3.$$

It is easy to see that $N_{\mathrm{VB}}^{\triangle}(L)\ (\mathrm{mod}\ 6)$ is 0/3 for $3p \pm 1$ odd/even, which is independent of $q$. Therefore, the topology of this triangular array inherited from $L_1 = (3p \pm 1) \times \sqrt{3}$ agrees with the diagnose based on $N_{\mathrm{VB}}^{\triangle}\ (\mathrm{mod}\ 6)$ as it should be.

**Table S1.** Topology, parity index and number of valence bands for graphene with $m\sqrt{3} \times m\sqrt{3}$ nanohole array. ✓/×: topological/trivial; $N_\Gamma^+/N_\mathrm{M}^+$: number of valence states with even parity at $\Gamma/\mathrm{M}$ point; $N_{\mathrm{VB}}^{\triangle}/N_{\mathrm{VB}}^{\hexagon}$: number of valence bands in triangular/honeycomb nanohole array. N/A: not applicable as the structure does not exist.

| $m$ | Triangular | $(N_\Gamma^+, N_\mathrm{M}^+)$ | $N_{\mathrm{VB}}^{\triangle}$ | Honeycomb | $(N_\Gamma^+, N_\mathrm{M}^+)$ | $N_{\mathrm{VB}}^{\hexagon}$ |
|---|---|---|---|---|---|---|
| 2 | ✓ | (3, 5) | 9 | N/A | N/A | N/A |
| 3 | × | (12, 12) | 24 | × | (10, 10) | 21 |
| 4 | ✓ | (21, 23) | 45 | × | (21, 21) | 42 |
| 5 | × | (36, 36) | 72 | ✓ | (36, 34) | 69 |
| 6 | ✓ | (51, 53) | 105 | ✓ | (49, 51) | 102 |
| 7 | × | (72, 72) | 144 | ✓ | (72, 70) | 141 |
| 8 | ✓ | (93, 95) | 189 | × | (93, 93) | 186 |
| 9 | × | (120, 120) | 240 | × | (118, 118) | 237 |
| 10 | ✓ | (147, 149) | 297 | × | (147, 147) | 294 |
| 11 | × | (180, 180) | 360 | ✓ | (180, 178) | 357 |

**Table S2.** Topology, parity index and number of valence bands for graphene with $3m \times 3m$ nanohole array.

| $m$ | Triangular | $(N_\Gamma^+, N_\mathrm{M}^+)$ | $N_{\mathrm{VB}}^{\triangle}$ | Honeycomb | $(N_\Gamma^+, N_\mathrm{M}^+)$ | $N_{\mathrm{VB}}^{\hexagon}$ |
|---|---|---|---|---|---|---|
| 1 | × | (3, 3) | 6 | N/A | N/A | N/A |
| 2 | ✓ | (15, 17) | 33 | ✓ | (13, 15) | 30 |
| 3 | × | (39, 39) | 78 | × | (37, 37) | 75 |
| 4 | ✓ | (69, 71) | 141 | ✓ | (71, 69) | 138 |

**Table S3.** Topology, parity index and number of valence bands for graphene with $4\sqrt{3} \times 4\sqrt{3}$ triangular array of nanoholes with increasing sizes. Size: number of removed carbon atoms per hole.

| Size | Topology | $(N_{\Gamma}^{+}, N_{M}^{+})$ | $N_{VB}^{\triangle}$ | $N_{VB}^{\triangle}$ (mod 6) |
|---|---|---|---|---|
| 6 | ✓ | (21, 23) | 45 | 3 |
| 12 | × | (21, 21) | 42 | 0 |
| 24 | × | (18, 18) | 36 | 0 |
| 36 | × | (15, 15) | 30 | 0 |
| 42 | ✓ | (13, 15) | 27 | 3 |
| 54 | ✓ | (9, 11) | 21 | 3 |

**Table S4.** Topology, parity index and number of valence bands for graphene with $4\sqrt{3} \times 4\sqrt{3}$ honeycomb array of nanoholes with increasing sizes.

| Size | Topology | $(N_{\Gamma}^{+}, N_{M}^{+})$ | $N_{VB}^{\hexagon}$ | $N_{VB}^{\hexagon}$ (mod 6) |
|---|---|---|---|---|
| 6 | × | (21, 23) | 42 | 0 |
| 12 | ✓ | (16, 18) | 36 | 0 |

**Supplementary Note II:** Energy gapping opening at the $\Gamma$ point of folded BZ of superstructures with $C_{6v}$ symmetry and eigen functions have been discussed previously, see for example Ref. [58] in the maintext. An effective Hamiltonian around the $\Gamma$ point of the superstructure is given

$$H(\boldsymbol{k} \to \Gamma) = M\tau_x + v\tau_z(k_x\sigma_x - k_y\sigma_y)$$

on the basis of $(|+, \mathrm{K_{Gr}}\rangle, |-, \mathrm{K_{Gr}}\rangle, |+, \mathrm{K'_{Gr}}\rangle, |-, \mathrm{K'_{Gr}}\rangle)$ with $\pm$ representing orbital-angular-momentum (OAM) eigenvalues $e^{\pm i2\pi/3}$ of $C_3$ symmetry, where the Pauli matrix $\tau/\sigma$ refers to the valley/OAM degrees of freedom. Note that the basis is defined in the unit cell of the superstructure, whereas the wave functions are formed in terms of those at $\mathrm{K_{Gr}}$ and $\mathrm{K'_{Gr}}$ of the pristine graphene. In the above effective Hamiltonian, the intervalley couplings $M\tau_x$ induced by nanohole array and/or non-uniform hopping integrals open an energy gap in the otherwise Dirac cones, and mix the wave functions with the same OAM eigenvalues of $C_3$ symmetry to form those with the OAM eigenvalues of $C_6$ symmetry:

$$|p_+\rangle = \frac{1}{\sqrt{2}}(|+, \mathrm{K_{Gr}}\rangle + |+, \mathrm{K'_{Gr}}\rangle),$$
$$|d_-\rangle = \frac{1}{\sqrt{2}}(|+, \mathrm{K_{Gr}}\rangle - |+, \mathrm{K'_{Gr}}\rangle),$$
$$|p_-\rangle = \frac{1}{\sqrt{2}}(|-, \mathrm{K_{Gr}}\rangle + |-, \mathrm{K'_{Gr}}\rangle),$$
$$|d_+\rangle = \frac{1}{\sqrt{2}}(|-, \mathrm{K_{Gr}}\rangle - |-, \mathrm{K'_{Gr}}\rangle),$$

where $|p_\pm\rangle$ and $|d_\pm\rangle$ are eigenstates of OAM eigenvalues $e^{\pm i\pi/3}$ and $e^{\pm i2\pi/3}$ of $C_6$ symmetry, respectively. It is obvious that, at the $\Gamma$ point, the time-reversal symmetry guarantees the degeneracy between $|p_\pm\rangle$ and between $|d_\pm\rangle$, with an energy gap $2M$ between $p$ and $d$ modes.